\documentclass[cameraready]{Interspeech}

\usepackage{cite}
\usepackage{nicefrac}
\usepackage{caption}
\usepackage{pifont}
\usepackage{multirow}
\usepackage{hhline}
\usepackage{graphicx}
\usepackage{hyperref}
\usepackage{booktabs}
\usepackage{amsmath}

\title{ReDimNet2: Scaling Speaker Verification via Time-Pooled Dimension Reshaping}

\author[affiliation={1}]{Ivan}{Yakovlev}
\author{Anton}{Okhotnikov}

\address{
    $^1$ Palabra AI, London, United Kingdom
}
\email{i.yakovlev@palabra.ai, ant.okhotnikov@gmail.com}


\newcommand{\ours}[1]{\textbf{#1}}

\begin{document}

\maketitle

\begin{abstract}
We present ReDimNet2, an improved neural network architecture for extracting utterance-level speaker representations that builds upon the ReDimNet dimension-reshaping framework. The key modification in ReDimNet2 is the introduction of pooling over the time dimension within the 1D processing pathway. This operation preserves the nature of the 1D feature space, since 1D features remain a reshaped version of 2D features regardless of temporal resolution, while enabling significantly more aggressive scaling of the channel dimension without proportional compute increase. We introduce a family of seven model configurations (B0-B6) ranging from 1.1M to 12.3M parameters and 0.33 to 13 GMACs. Experimental results on VoxCeleb1 benchmarks demonstrate that ReDimNet2 improves the Pareto front of computational cost versus accuracy at every scale point compared to ReDimNet, achieving 0.287\% EER on Vox1-O with 12.3M parameters and 13 GMACs.
\end{abstract}

\section{Introduction}

Speaker recognition is a specialized field aiming at identifying or verifying individuals through their distinct voice features. Deep neural networks have emerged as the dominant technology for extracting speaker embeddings used for Speaker Verification (SV) and related tasks. Extensive research has advanced the field through new datasets \cite{Vox2, VoxBlink, 3DSpeaker, VoxTube}, architecture design \cite{xvector, desplanques2020ecapa, next_tdnn, liu2022mfa, pcf_tdnn, dtdnn, Cam, zhang2022mfaconformer, thienpondt2024ecapa2, garcia2020magneto, resnext, eres2net, dfresnet, gemini}, and loss functions \cite{deng2019arcface, SF2}.

A variety of architectures have emerged including 1D \cite{xvector, desplanques2020ecapa, dtdnn, pcf_tdnn, next_tdnn} and 2D \cite{garcia2020magneto, resnext, eres2net, dfresnet, gemini} CNNs, their hybrids \cite{liu2022mfa, Cam, thienpondt2024ecapa2}, and self-attention networks \cite{zhang2022mfaconformer}. Each approach has distinct advantages: 1D models offer efficiency and direct temporal analysis, 2D architectures provide frequency translational invariance \cite{thienpondt2021integrating}, and hybrid systems aim to combine both. More recently, large self-supervised models such as WavLM \cite{wavlm} and W2V-BERT 2.0 \cite{w2vbert2, w2vbert2_sv} have achieved strong results but at significantly higher computational costs.

ReDimNet \cite{redimnet_v1} introduced a novel approach based on dimensionality reshaping between 2D and 1D representations, enabling seamless integration of both block types with residual connections throughout the network. Its compact and accurate speaker embeddings have since been adopted in diverse downstream tasks: speaker conditioning for zero-shot text-to-speech synthesis \cite{voxtream}, speaker similarity evaluation in speech enhancement and speech generation \cite{gsepf, pfluxtts}, and as a compact speaker encoder for personal voice activity detection \cite{pvad_lin25}. While ReDimNet achieved state-of-the-art results, its design constraint of preserving time resolution throughout the network limited the scalability of the channel dimension: increasing channels without time reduction leads to quadratic growth in computation within the 1D pathway.

\begin{figure}[!t]
  \centering
  \includegraphics[width=\columnwidth]{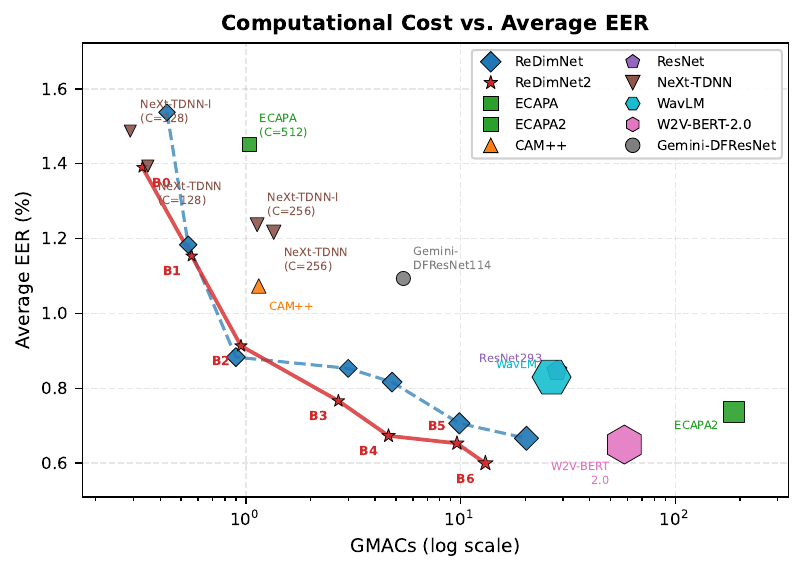}
  \caption{{\bfseries Computational Cost vs.\ Average EER.} EER is averaged over Vox1-O, Vox1-E, and Vox1-H protocols. Marker area is proportional to parameter count, color indicates model family. Dashed line: ReDimNet, solid line: ReDimNet2. GMACs measured on 2-second input.}
  \label{fig:cost_vs_eer}
\vspace{-10pt}
\end{figure}

In this paper, we present ReDimNet2, which addresses this limitation by introducing pooling over the time dimension in the 1D pathway. The key insight is that time-pooling does not fundamentally alter the 1D feature space - 1D features remain a reshaped version of 2D features, so the residual connection and dimension-reshape logic remain valid despite the reduced temporal resolution. This simple modification allows more aggressive channel scaling and yields consistently improved accuracy at every compute budget (Fig.~\ref{fig:cost_vs_eer}).

\section{Model Architecture}

\begin{figure*}[!t]
  \centering
  \includegraphics[width=\textwidth]{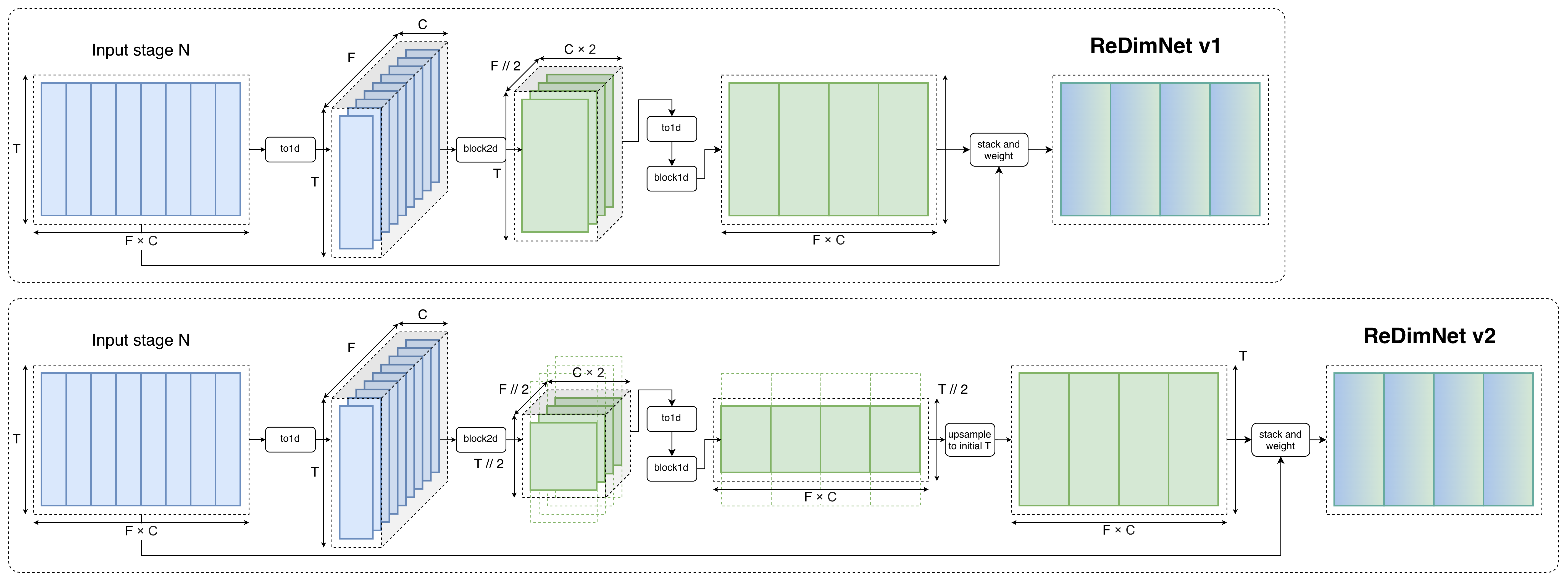}
  \caption{{\bfseries ReDimNet2 architecture overview} for a single stage. \textit{Top:} ReDimNet v1 processes 1d input through a \texttt{to2d} reshape operation, \texttt{block2d} (frequency stride only), a \texttt{to1d} reshape operation followed by a \texttt{block1d}, and stage-wise weighted aggregation (``stack and weight''), preserving the time dimension $T$ throughout. \textit{Bottom:} ReDimNet2 applies an additional stride over the time axis inside the 2D block, halving both $F$ and $T$. The 1D block then operates on the shorter sequence ($T/2$), and an upsample-to-initial-$T$ step restores the original temporal resolution before aggregation.}
  \label{fig:arch_scheme}
\end{figure*}

\subsection{ReDimNet recap}

ReDimNet \cite{redimnet_v1} is built around dimensionality reshaping between 2D and 1D feature maps. The architecture constrains all 2D feature maps to have a constant volume $V = C \cdot F \cdot T$, where $C$, $F$, $T$ are channels, frequency bins, and time steps, respectively. This is achieved by (1)~eliminating all strides over the time axis, and (2)~synchronizing frequency strides with channel growth rates across stages. Given constant volume, any 2D feature map $(C_i, F_i, T)$ can be reshaped to a common 1D representation $(C_i \cdot F_i, T) = (C_0 \cdot F_0, T)$, enabling residual connections between arbitrary stages. Joint 1D+2D blocks process features through interleaved 2D convolutional subblocks (basic ResNet blocks) and 1D time-contextual subblocks (ConvNeXt-like 1D convolutions combined with multi-head attention), with Attentive Statistics Pooling \cite{att_pool} for utterance-level aggregation. We reuse the best-performing 1D and 2D block types identified through ablation studies in \cite{redimnet_v1}.

\subsection{ReDimNet2: time-pooled dimension reshaping}

The central modification in ReDimNet2 is the introduction of \textbf{pooling over the time dimension} within the 1D processing pathway (Fig.~\ref{fig:arch_scheme}). In the original ReDimNet, the time axis $T$ is preserved at the input resolution throughout the entire network. While this preserves full temporal information, the computational cost of the 1D subblocks scales linearly with both channels $C$ and time steps $T$, making large-$C$ configurations increasingly expensive.

\textbf{Pooling mechanism.} ReDimNet2 applies time-pooling at intermediate stages using the same strided conv2d layer that implements frequency downsampling in the original architecture - the one that preserves spatial volume by doubling channels as $F$ is halved. When applied in the time direction, the channel dimension is not adjusted, which softly relaxes the constant-volume constraint $C \cdot F \cdot T$. Table~\ref{feat-map-size-table} shows the resulting feature map shapes for the B3 configuration. Frequency-pool stages (2, 4, 6) preserve the volume exactly, while time-pool stages (3, 5) reduce $T$ by half without changing $C$ or $F$.

\begingroup
\begin{table}[!t]
\centering
\caption{{\bfseries Feature map sizes for B3 v2 configuration.} $S_f$, $S_t$: frequency and time strides. Time-pool stages (3, 5) halve $T$ without adjusting $C$, softly relaxing the $C{\cdot}F{\cdot}T$ constraint.}
\label{feat-map-size-table}
\scalebox{0.75}{
\begin{tabular}{c|c|c|c|c|c|c}
Block \# & In shape & $S_{f}$ & $S_{t}$ & Channels & Out shape & Volume \\
\hline
1 & $(C,\;F,\;T)$ & 1 & 1 & $C$ & $(C,\;F,\;T)$ & $C{\cdot}F{\cdot}T$ \\
2 & $(C,\;F,\;T)$ & 2 & 1 & $2C$ & $(2C,\;F/2,\;T)$ & $C{\cdot}F{\cdot}T$ \\
3 & $(2C,\;F/2,\;T)$ & 1 & 2 & $2C$ & $(2C,\;F/2,\;T/2)$ & $C{\cdot}F{\cdot}T/2$ \\
4 & $(2C,\;F/2,\;T/2)$ & 2 & 1 & $4C$ & $(4C,\;F/4,\;T/2)$ & $C{\cdot}F{\cdot}T/2$ \\
5 & $(4C,\;F/4,\;T/2)$ & 1 & 2 & $4C$ & $(4C,\;F/4,\;T/4)$ & $C{\cdot}F{\cdot}T/4$ \\
6 & $(4C,\;F/4,\;T/4)$ & 2 & 1 & $8C$ & $(8C,\;F/8,\;T/4)$ & $C{\cdot}F{\cdot}T/4$ \\
\hline
\end{tabular}
}
\end{table}
\endgroup

\textbf{Residual connections.} In the original ReDimNet, all stage outputs share the same 1D shape $(B, C{\cdot}F, T)$ and can be directly aggregated via learned stage-wise weights. In ReDimNet2, stages produce 1D feature maps at different temporal lengths: $(B, C{\cdot}F, T)$, $(B, C{\cdot}F, T/2)$, $(B, C{\cdot}F, T/4)$, and so on. To maintain residual connectivity, we apply nearest-neighbor upsampling to all feature maps prior to the stage-wise weighted aggregation, aligning them to the input temporal resolution $T^*$:
\begin{equation*}
\begin{aligned}
&[(B,\,C{\cdot}F,\,T),\;(B,\,C{\cdot}F,\,\tfrac{T}{2}),\;(B,\,C{\cdot}F,\,\tfrac{T}{4}),\;\ldots]\\
&\;\to\;
[(B,\,C{\cdot}F,\,T^*),\;(B,\,C{\cdot}F,\,T^*),\;\ldots]
\end{aligned}
\end{equation*}
where $T^* = T$. Each stage still processes its own features at reduced temporal resolution; upsampling is applied only at the aggregation point, so the compute savings within each stage are preserved.

\textbf{Dual efficiency benefit.} Time-pooling benefits both processing pathways. The 1D subblocks benefit directly from shorter sequences, reducing their cost proportionally. The 2D subblocks benefit even more: since the 1D-to-2D reshape produces 2D feature maps whose spatial extent depends on the sequence length, reduced $T$ compresses the 2D representations as well, compounding the savings. This dual reduction enables more aggressive scaling of the channel dimension $C$ within the same GMACs budget.

The freed compute budget is reallocated to wider models (higher $C$), which empirically yields better speaker discriminability at matched computational cost.

\subsection{Model configurations}

Following \cite{tan2020efficientnet}, we define seven configurations B0--B6 bounded by computational complexity in GMACs:
\textit{B0 $<$ 0.5, B1 $\approx$ 0.5, B2 $<$ 1.5, B3 $<$ 3.0, B4 $<$ 5.5, B5 $<$ 10.0, B6 $>$ 10.0.}
ReDimNet2 scales from 1.1M params / 0.33 GMACs at B0 to 12.3M params / 13.0 GMACs at B6.

\begingroup
\newcommand{\sr}{\rule[-0.25cm]{0pt}{0.7cm}}
\begin{table*}[!t]
\centering
\caption{{\bfseries Evaluation results on VoxCeleb1 cleaned protocols (EER, \%).} All models trained on VoxCeleb2-dev unless noted. GMACs measured on 2-second segments. * = estimated values. Models from WeSpeaker or ECAPA2 repositories retested in our environment. $^\dagger$ = trained on VoxBlink2 + VoxCeleb2.}
\label{tab:main-comparison}
\setlength\tabcolsep{5pt}
\renewcommand{\arraystretch}{0.95}
\resizebox{0.85\textwidth}{!}{
\begin{tabular}{llccccc}
\toprule
  \textbf{Model} &
  \textbf{Params} &
  \textbf{GMACs} &
  \textbf{LM} &
  \textbf{Vox1-O} &
  \textbf{Vox1-E} &
  \textbf{Vox1-H} \\
\midrule
ReDimNet-B0 \cite{redimnet_v1}          & 1.0M  & 0.43  & \ding{51} & 1.16  & 1.25  & 2.20  \\
\ours{ReDimNet2-B0}                       & \ours{1.1M}  & \ours{0.33}  & \ding{51} & \ours{1.04}  & \ours{1.16}  & \ours{1.97}  \\
\midrule
NeXt-TDNN-l (C=128) \cite{next_tdnn}    & \ours{1.6M}  & \ours{0.29*} & \ding{55} & 1.10  & 1.24  & 2.12  \\
NeXt-TDNN (C=128) \cite{next_tdnn}      & 1.9M  & 0.35* & \ding{55} & 1.03  & 1.17  & 1.98  \\
ReDimNet-B1 \cite{redimnet_v1}          & 2.2M  & 0.54  & \ding{51} & 0.85  & 0.97  & 1.73  \\
\ours{ReDimNet2-B1}                       & 2.1M  & 0.56  & \ding{51} & \ours{0.78}  & \ours{0.96}  & \ours{1.72}  \\
\midrule
ECAPA (C=512) \cite{desplanques2020ecapa, dfresnet}  & 6.2M & 1.04 & \ding{55} & 0.94 & 1.21 & 2.20  \\
NeXt-TDNN-l (C=256) \cite{next_tdnn}    & 6.0M  & 1.13* & \ding{55} & 0.81  & 1.04  & 1.86  \\
CAM++ \cite{Cam, wang2023wespeaker}      & 7.2M  & 1.15  & \ding{51} & 0.71  & 0.85  & 1.66  \\
NeXt-TDNN (C=256) \cite{next_tdnn}      & 7.1M  & 1.35* & \ding{55} & 0.79  & 1.04  & 1.82  \\
ReDimNet-B2 \cite{redimnet_v1}          & 4.7M  & \ours{0.90}  & \ding{51} & \ours{0.57}  & \ours{0.76}  & \ours{1.32}  \\
\ours{ReDimNet2-B2}                       & \ours{3.6M}  & 0.95  & \ding{51} & \ours{0.57}  & \ours{0.76}  & 1.41  \\
\midrule
ReDimNet-B3 \cite{redimnet_v1}          & \ours{3.0M}  & 3.00  & \ding{51} & 0.50  & 0.73  & 1.33  \\
\ours{ReDimNet2-B3}                       & 4.1M  & \ours{2.70}  & \ding{51} & \ours{0.42}  & \ours{0.66}  & \ours{1.22}  \\
\midrule
Gemini-DFResNet114 \cite{gemini}         & 6.5M  & 5.42  & \ding{55} & 0.77  & 0.91  & 1.60  \\
ReDimNet-B4 \cite{redimnet_v1}          & \ours{6.3M}  & 4.80  & \ding{51} & 0.51  & 0.68  & 1.26  \\
\ours{ReDimNet2-B4}                       & 6.6M  & \ours{4.62}  & \ding{51} & \ours{0.37}  & \ours{0.58}  & \ours{1.07}  \\
\midrule
ReDimNet-B5 \cite{redimnet_v1}          & 9.2M  & 9.87  & \ding{51} & 0.43  & 0.61  & 1.08  \\
\ours{ReDimNet2-B5}                       & \ours{8.9M}  & \ours{9.62}  & \ding{51} & \ours{0.33}  & \ours{0.56}  & \ours{1.07}  \\
\midrule
ResNet293 \cite{wang2023wespeaker, resnet}  & 28.6M & 28.10 & \ding{51} & 0.53 & 0.71 & 1.30  \\
ECAPA2 \cite{thienpondt2024ecapa2}       & 27.1M & 187.0*& \ding{51} & 0.44  & 0.62  & 1.15  \\
WavLM \cite{wavlm}                       & 324M  & 26.53 & \ding{51} & 0.52  & 0.63  & 1.34  \\
W2V-BERT 2.0 \cite{w2vbert2_sv}          & 587M  & 57.90 & \ding{51} & 0.38  & \ours{0.51}  & 1.06  \\
ReDimNet-B6 \cite{redimnet_v1}          & 15.0M & 20.27 & \ding{51} & 0.40  & 0.55  & 1.05  \\
\ours{ReDimNet2-B6}                       & \ours{12.3M} & \ours{13.05} & \ding{51} & \ours{0.29}  & 0.52  & \ours{0.99}  \\
\midrule
\multicolumn{7}{l}{\textit{Models trained on extended data (VoxBlink2 + VoxCeleb2):}} \\
SimAM-ResNet34$^\dagger$ \cite{simam_resnet}  & 25.2M & 18.20 & - & 0.42 & 0.62 & 1.12  \\
SimAM-ResNet100$^\dagger$ \cite{simam_resnet} & 50.2M & 57.24 & - & 0.23 & 0.46 & 0.87  \\
W2V-BERT 2.0$^\dagger$ \cite{w2vbert2_sv}    & 587M  & 57.90 & - & 0.14 & 0.31 & 0.73  \\
\bottomrule
\end{tabular}
}
\end{table*}
\endgroup

\section{Experimental Setup}

We train ReDimNet2 on the VoxCeleb2 \cite{Vox2} using a two-stage training approach with the \texttt{wespeaker} \cite{wang2023wespeaker} training pipeline.

\subsection{Pretraining stage}
\vspace{-1.5pt}
Models are optimized using SGD with Nesterov momentum ($m=0.9$) and weight decay of $2{\times}10^{-5}$. We use 80-dimensional mean-normalized Mel filter bank log-energies with a 25\,ms frame length and 10\,ms shift. During pretraining, 2-second segments are randomly cropped, and augmentations from the MUSAN \cite{snyder2015musan} and RIR \cite{szoke2019building} datasets are applied following \cite{garcia2020magneto}. Two-fold speed perturbation with factors 0.9 and 1.1 is used. SphereFace2-C (SF2-C) \cite{SF2} is used as the default loss function, with the margin scheduled from 0.0 to 0.2 over epochs 20--40, then held constant. Learning rate follows an exponential decay schedule with 6-epoch warmup, $lr_{\max}=0.1$, $lr_{\min}=6{\times}10^{-5}$. Each model configuration was trained with different GPU setup for maximum resources utilization. We found that multi-GPU training setup (8 x H100) gives better results for larger models due to the linear scaling of effective learning rate induced by large effective batch size. 


\subsection{Large-Margin Finetuning stage}

For the LM finetuning stage \cite{LM}, training utterance length is expanded to 6 seconds, speed perturbations are turned off, and the SF2-C margin is set to a constant 0.3. Learning rate is reduced to $10^{-4}$ with exponential decay. This stage runs for 5 epochs.

\subsection{Evaluation}

Performance is assessed using cleaned protocols of VoxCeleb1 \cite{Nagrani19}: Vox1-O (Original), Vox1-E (Extended), and Vox1-H (Hard), employing the Equal Error Rate (EER). Models are scored with cosine similarity using full-length test utterances without score normalization. GMACs are computed using the \texttt{thop} library with a 2-second raw waveform input at 16\,kHz.

\begin{table}[!t]
\centering
\caption{{\bfseries Out-of-domain evaluation (EER, \%).} All models trained on VoxCeleb2-dev. Baselines from \cite{redimnet_v1}.}
\label{tab:ood}
\setlength\tabcolsep{4pt}
\scalebox{0.88}{
\begin{tabular}{lcccc}
\toprule
\textbf{Model} & \textbf{SITW} & \textbf{VOiCES} & \textbf{Vox1-B} & \textbf{Avg} \\
\midrule
CAM++ \cite{Cam}                          & 1.34 & 6.30  & 2.79 & 3.48 \\
ECAPA (C=1024) \cite{desplanques2020ecapa}& 1.67 & 5.31  & 3.48 & 3.49 \\
ResNet293 \cite{resnet}                   & 1.67 & 5.14  & 2.23 & 3.01 \\
ECAPA2 \cite{thienpondt2024ecapa2}        & 3.64 & 13.26 & 1.81 & 6.24 \\
ReDimNet-B6 \cite{redimnet_v1}            & 0.77 & 3.19  & 1.66 & 1.87 \\
\ours{ReDimNet2-B6}                       & \ours{0.75} & \ours{3.13} & \ours{1.63} & \ours{1.84} \\
\bottomrule
\end{tabular}
}
\end{table}

\section{Results}

Evaluation results for all ReDimNet2 configurations and baselines are presented in Table~\ref{tab:main-comparison} and visualized in Fig.~\ref{fig:cost_vs_eer}. The B0--B6 scaling comparison constitutes the primary evidence for the effectiveness of time-pooling: almost at every matched compute budget, ReDimNet2 achieves lower EER than the corresponding ReDimNet configuration. This seven-point comparison across the full Pareto front provides strong evidence since GMACs are the predominant factor governing accuracy in speaker verification. Removing all time strides from ReDimNet2-B3 (2.7 GMACs) would increase its compute to approximately 4.0 GMACs, directly comparable to ReDimNet-B4 at 4.8 GMACs - a comparison point present in Table~\ref{tab:main-comparison}. As all B0--B6 pairs are aligned in GMACs and parameter count, the seven paired-budget comparisons also constitute a complete cross-scale ablation.

The gains are most pronounced at larger scales, where time-pooling enables more aggressive channel scaling. ReDimNet2-B6 achieves \textbf{0.29\%} EER on Vox1-O - a 28\% relative improvement over ReDimNet-B6 - while requiring 36\% fewer GMACs and 18\% fewer parameters. Even at the smallest scale, ReDimNet2-B0 improves from 1.16\% to 1.04\% EER on Vox1-O, confirming that the benefit is consistent across all model sizes. Moreover, ReDimNet2-B6 outperforms WavLM (324M params) and approaches W2V-BERT~2.0 (587M params) while being 48$\times$ smaller in parameter count. Mid-range configurations are equally notable: ReDimNet2-B3 surpasses ECAPA2 on Vox1-O at 69$\times$ fewer GMACs, and every configuration from B3 onwards exceeds ResNet293 (28.6M params).

\textbf{Out-of-domain generalization.} We also evaluated ReDimNet2-B6 on out-of-domain test sets following the protocol used in \cite{redimnet_v1}: the SITW core-core set \cite{mclaren2016speakers}, the VOiCES evaluation set \cite{nandwana2019voices}, and the VoxCeleb1-B hard protocol \cite{nam2023disentangled}. Results are shown in Table~\ref{tab:ood}. ReDimNet2-B6 achieves lower EER than ReDimNet-B6 on all three out-of-domain sets, demonstrating that time-pooling does not hurt generalization.

\textbf{Training stability.} We observed that results for smaller models (B0--B3) are stable across random seeds, with mean$\pm$std EER of $0.42\pm0.01$\% for B3 on Vox1-O. However, for larger models (B4--B6), we noticed increased variability: B6 achieves $0.32\pm0.05$\% EER on Vox1-O across multiple runs. This suggests that while the proposed architecture scales well, training larger models may benefit from additional regularization or careful hyperparameter tuning to ensure stability.

\section{Conclusions}

In this paper we presented ReDimNet2, an improved speaker embedding architecture that extends ReDimNet with time-pooling in the 1D processing pathway. We showed that time-pooling is compatible with the dimension-reshaping framework: the same volume-preserving conv2d layer used for frequency downsampling handles time pooling as well, and nearest-neighbor upsampling at the aggregation point maintains full residual connectivity. The resulting model family (B0--B6) achieves a better accuracy-efficiency trade-off than ReDimNet at every matched compute budget. In particular, ReDimNet2-B6 reaches 0.29\% EER on Vox1-O and 0.99\% on Vox1-H with only 12.3M parameters and 13 GMACs, while also preserving strong out-of-domain generalization. We conclude that time-pooling is a simple yet effective strategy for scaling dimension-reshaping architectures for speaker verification. Model code, training recipes, and pretrained weights are released at \mbox{\url{https://github.com/PalabraAI/redimnet2}}.

\section{Disclosure of AI Tool Use}

A generative AI assistant (Claude, Anthropic) was used for editing and polishing the manuscript text. All scientific content, experimental design, model architecture, and analysis were produced entirely by the authors.

\bibliographystyle{IEEEtran}
\bibliography{mybib}

\end{document}